\documentclass[%
superscriptaddress,
preprint,
amsmath,amssymb,
]{revtex4-1}

\usepackage{graphicx}
\usepackage{dcolumn}
\usepackage{bm}

\usepackage{amsthm}
\usepackage{amsmath}
\usepackage{amssymb}
\usepackage{graphicx}
\usepackage{url}
\usepackage{float}
\usepackage{bm}
\usepackage{ifthen}
\usepackage[usenames,dvipsnames]{color}
\usepackage{mathrsfs}
\usepackage[colorlinks=true,citecolor=blue,urlcolor=black]{hyperref}
\usepackage{float}
\usepackage{subfigure}
\usepackage{enumitem}
\usepackage{color}
\usepackage{setspace}
\usepackage{braket}

\usepackage{lipsum}

\newcommand{\be}{\begin{equation}}
	\newcommand{\ee}{\end{equation}}

\begin{document}

\title{Pre-fixed Threshold Real Time Selection Method in Free-space Quantum Key Distribution}

\author{Wenyuan Wang}
\affiliation{Centre for Quantum Information and Quantum Control (CQIQC), Dept. of Electrical \& Computer Engineering and Dept. of Physics, University of Toronto, Toronto,  Ontario, M5S 3G4, Canada}

\author{Feihu Xu}
\affiliation{Shanghai Branch, National Laboratory for Physical Sciences at Microscale, University of Science and Technology of China, Shanghai, 201315, China}

\author{Hoi-Kwong Lo}
\affiliation{Centre for Quantum Information and Quantum Control (CQIQC), Dept. of Electrical \& Computer Engineering and Dept. of Physics, University of Toronto, Toronto,  Ontario, M5S 3G4, Canada}

\begin{abstract}
	Free-space Quantum key distribution (QKD) allows two parties to share a random key with unconditional security, between ground stations, between mobile platforms, and even in satellite-ground quantum communications. Atmospheric turbulence causes fluctuations in transmittance, which further affect the quantum bit error rate (QBER) and the secure key rate. Previous post-selection methods to combat atmospheric turbulence require a threshold value determined after all quantum transmission. In contrast, here we propose a new method where we pre-determine the optimal threshold value even before quantum transmission. Therefore, the receiver can discard useless data immediately, thus greatly reducing data storage requirement and computing resource. Furthermore, our method can be applied to a variety of protocols, including, for example, not only single-photon BB84, but also asymptotic and finite-size decoy-state BB84, which can greatly increase its practicality.
\end{abstract}

\date{\today}
\maketitle


\section{Introduction}

Quantum key distribution (QKD), first proposed by Bennett and Brassard\cite{bb84} in 1984, allows two parties to securely share a random secret key, which can be further combined with cryptographic protocols, such as one-time pad\cite{onetimepad}, to encode messages with unconditional security unachievable by classical cryptography. 

There has been increasing interest in implementing QKD through free-space channels. A major attraction for free-space QKD is that, performed efficiently, it could potentially be applied to airborne or maritime quantum communications where participating parties are on mobile platforms. Furthermore, it could even enable applications for ground to satellite quantum communications, and eventually, global quantum communication network.

Free-space quantum communication has seen great advances over the past 25 years. The first demonstration of free-space QKD was published by Bennett et al. from IBM research in 1992 \cite{freespace_IBM} over 32cm of free-space channel, which was also the first successful demonstration of experimental QKD. Over the next two decades, numerous demonstrations for free-space QKD have been made. In 1998, Buttler and Hughes et al. \cite{freespace_Hughes} have performed QKD over 1km of free-space channel outdoors at nighttime. In 2005, Peng et al. \cite{freespace_2005_Peng} performed distribution of entangled photons over 13km. In 2007, two successful experimental ground-to-ground free-space QKD experiments based on BB84 and E91 protocol \cite{free2007,freespace_144km_E91} were implemented over a 144km link between the Canary Island of La Palma and Tenerife. Ling et al. \cite{freespace_urban_E91} performed another entanglement-based QKD Experiment with modified E91 protocol over 1.4km in urban area in 2008. In 2012, Yin et al. and Ma et al. \cite{freespace_2012_100km, freespace_2012_143km} respectively performed quantum teleportation over 100km and 143 km.

In recent years, free-space QKD has also seen much development over rapidly moving platforms, with an air-to-ground experiment in 2013 by Nauerth et al. \cite{freespace_plane} over a plane 20km from ground, a demonstration of QKD with devices on moving turntable and floating hot ballon over 20km and 40km by J-Y Wang et al. \cite{freespace_2013_satellite} in 2013, a very recent report on drone-to-drone QKD experiment in 2017 by D. Hill et al. \cite{freespace_drone}, and notably, satellite-based quantum communication experiments in 2017 \cite{freespace_satellite1, freespace_satellite2, freespace_satellite3}, including a QKD experiment from a quantum satellite to the ground over a 1200km channel. Meanwhile, there is a lot of interest in doing QKD in a maritime environment either over sea water\cite{freespace_maritime_data} or through sea water \cite{freespace_underwater}. A study on quantum communication performance over a turbulent free-space maritime channel using real atmospheric data can be found in Ref.\cite{freespace_maritime_data}.

A major characteristic of a free-space channel is its time-dependent transmittance. This is due to the temporal fluctuations of local refractive index in the free-space channel, i.e. \textit{atmospheric turbulence}, which causes scintillation and beam wandering \cite{freespacethesis}, and result in fluctuations in the channel transmittance, which in turn affect QKD performance. Therefore, addressing turbulence is a major challenge for QKD over free-space. This fluctuation due to turbulence can be modeled as a probability distribution, called the Probability Distribution of Transmission Coefficient (PDTC). Hence the real time transmittance $\eta$ is a random time-dependent variable that can be described by the PDTC. 

\begin{table*}[t]
	\caption{Comparison of transmittance post-selection methods in QKD through turbulence channel}
	\begin{center}
		\begin{tabular}{cccc}			
			Method & Threshold choice & Model of signals & Sampling of transmittance\\
			\hline
			ARTS\cite{probetest} & post-determined & single-photon & secondary probe laser\\
			SNRF\cite{SNRF} & post-determined & single-photon & detector count (coincidence) rate \\
			P-RTS & pre-determined & universal & universal
			
		\end{tabular}
	\end{center}
\end{table*}

As free-space channels have time-varying transmittance due to turbulence, the QBER (and hence the secure key rate) for QKD changes with time. In previous literature discussing free-space QKD, such as \cite{free2007,Hughes}, the time variance of the channel is ignored, i.e. the secure key rate is calculated based on the time-average of channel transmittance. Having knowledge of the PDTC, however, Vallone et al. proposed a method named Adaptive Real-Time Selection (ARTS)\cite{probetest} that acquires information about real-time transmittance fluctuation due to turbulence, and makes use of this information to perform post-selection and improve the key rate of QKD.

However, ARTS method needs to "adaptively" choose an optimal threshold by performing numerical optimization after collecting all the data. A similar proposal by Erven et al. \cite{SNRF} called "signal-to-noise-ratio-filter (SNRF)" also discusses the idea of using a threshold to post-select high-transmittance periods, but uses the quantum data itself rather than a secondary classical channel. However, it needs to numerically optimize the threshold after collecting all experiment data, too. 

Here we ask the question, is scanning through all acquired data after experiment and finding such an "adaptive" threshold really necessary? The answer is in fact no. In this paper, we propose a new method called "pre-fixed threshold real-time selection (P-RTS)", and show the important observation that, for post-selection based on transmittance in a turbulent channel, the optimal post-selection threshold is independent of the channel, and can be directly calculated from experimental parameters of the devices beforehand - thus simplifying the implementation and enabling post-selection of signals in real time, which can also reduce the data storage requirements and computational resources in Bob's system. This is because, instead of having to wait until all data is collected to optimize the threshold, Bob can immediately discard the data that are obtained below the pre-fixed threshold and doesn't need to store all collected data. Moreover, he doesn't need to have a model for the PDTC of the channel, and no longer need to run a numerical optimization to find optimal threshold, thus we can additionally save software development effort and computing resource for Bob, too.

Furthermore, both ARTS and SNRF are limited to single photon model only, while decoy-state must be used for QKD with practical weak coherent pulse (WCP) source. Here we also propose an universal framework for QKD through a channel with fluctuating transmittance, for not only single-photon BB84, but also practical decoy-state BB84 with WCP source, and decoy-state BB84 with finite-size effects (both of which we are the first to apply threshold post-selection to), thus greatly improving its usefulness in practice. We also propose a model to estimate the maximum improvement in key rate from using threshold post-selection, and show that with P-RTS method we can achieve a key rate very close to the maximum performance with an optimal threshold.

A comparison of P-RTS with post-selection methods in previous literature can be seen in Table I. As shown here, P-RTS has the great advantage of being able to predict the optimal threshold independently of the channel. This means that one no longer needs to store all the data after experiment and optimize the threshold, but can perform real-time selection with a single threshold, regardless of the actual channel turbulence and loss condition. Moreover, our result is valid not only for BB84 with single photons, but for any general protocol that has a fluctuating transmittance. It is also not restricted to transmittance sampling with a secondary laser as in ARTS, but for instance can also use observed photon count rates in a given time interval as in SNRF.

Lastly, we have performed a computer simulation to show the actual advantage of using P-RTS in practical decoy-state BB84, with up to 170\% improvement in decoy-state BB84 key rate for certain scenarios, or 5.1dB increase in maximum tolerant loss at $R=10^{-7}$, under medium-level turbulence. We also include a numerical demonstration for applying P-RTS to BB84 with finite-size effects, which still shows significant increase in rate even when total number of signals is limited, e.g. maximum tolerant loss at $R=10^{-7}$ gains an increase of 1.4dB to 5.2dB, for $N=10^{11}-10^{13}$ under high turbulence.

The organization of the paper is listed as follows: in section 2 we first present a brief recapitulation of ARTS method, and proceed to propose a universal framework for QKD key rate in turbulent channel. We then propose P-RTS method, and discuss how and why an optimal threshold can be pre-fixed, and show an upper bound for the rate of P-RTS. We also present numerical results from simulations to show how P-RTS behaves compared to no post-selection. Lastly, we discuss P-RTS in decoy-state BB84, for the asymptotic case in Section III and for finite-size regime in Section IV, and show with simulation results that P-RTS works effectively for both of them, too.

\section{Methods}
\subsection{The ARTS Method}

In Ref.\cite{probetheory}, Capraro et al performed an experiment to study the impact of turbulence on a quantum laser transmitted through an 143km channel on Canary Islands, and proposed the idea of improving SNR with a threshold at the expense of number of signals. Subsequently, in Ref.\cite{probetest}, Vallone et al from the same group performed an experiment of free-space single-photon B92 QKD through the same channel, and showed the effectiveness of using real-time transmittance information in a turbulent channel to improve secure key rate, by performing post-selection on signals with a threshold, hence naming the method adaptive real-time selection (ARTS). 

\begin{figure}[b]
	\includegraphics[scale=0.28]{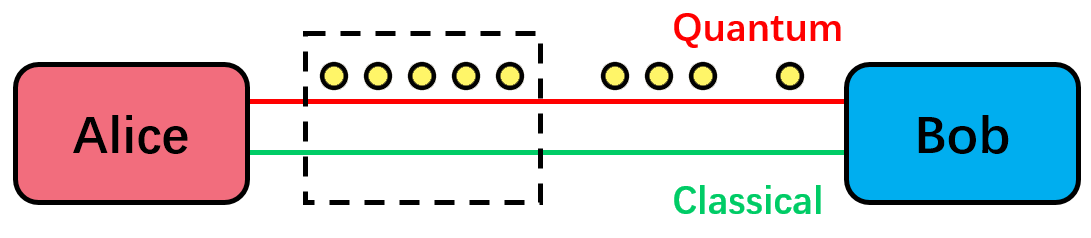}
	\caption{The ARTS setup by Vallone et al., where Alice and Bob are linked by a quantum channel and a classical channel. One can post-select quantum signals passing through the channel with high-transmittance, using a threshold on the corresponding classical channel signal}
	\label{fig:ARTS}
\end{figure}

This is realized by using a classical probe signal (a strong laser beam) alongside the quantum channel. In the quantum channel, the bits are polarization-encoded into quantum signals, which are detected by single-photon avalanche diodes (SPADs) that return click events. Meanwhile, the laser passing through the classical channel is detected by an avalanche photodetector that returns a voltage proportional to received light intensity, which is also proportional to the channel transmittance at that moment. An illustration of the setup can be seen in Fig. \ref{fig:ARTS}.

The key idea is that the transmittance of the classical channel will correspond to that of the quantum channel. Therefore, by reading voltage from the classical detector (defined as $V$), one can gain information of the periods of high transmittance. Combined with a threshold on the classical signal (defined as $V_T$), this information can be used to post-select only those quantum signals received by Bob during high transmittance periods (only when $V \geq V_T$), thus increasing the overall average transmittance, at the expense of a smaller number of signals due to post-selection. 

This post-selection increases the signal-to-noise ratio among the selected signals, and hence reduces the QBER, which subsequently increases the key rate. However, post-selection also reduces the total number of signals. Therefore, this becomes an optimization problem, and the choice of threshold value becomes critical. By numerically choosing an optimal threshold that maximizes the rate, it is possible to acquire a secure key rate much higher than before applying post-selection. This, as defined in Ref.\cite{probetest}, is called the adaptive real time selection (ARTS) method.

\subsection{Universal Framework for QKD Key Rate in a Turbulent Channel}

In this section, we will expand upon this threshold post-selection idea from ARTS, and apply it to a general framework of post-selection upon transmittance. We will then discuss the effects of threshold post-selection based on transmittance on the secure key rate. Our following discussions will be based on the channel transmittance $\eta$ only, and they are not limited to the secondary-laser transmittance sampling as in ARTS, but can be applied to any sampling method of transmittance, including photon count rate such as in SNRF.

\begin{figure}[b!]
	\includegraphics[scale=0.26]{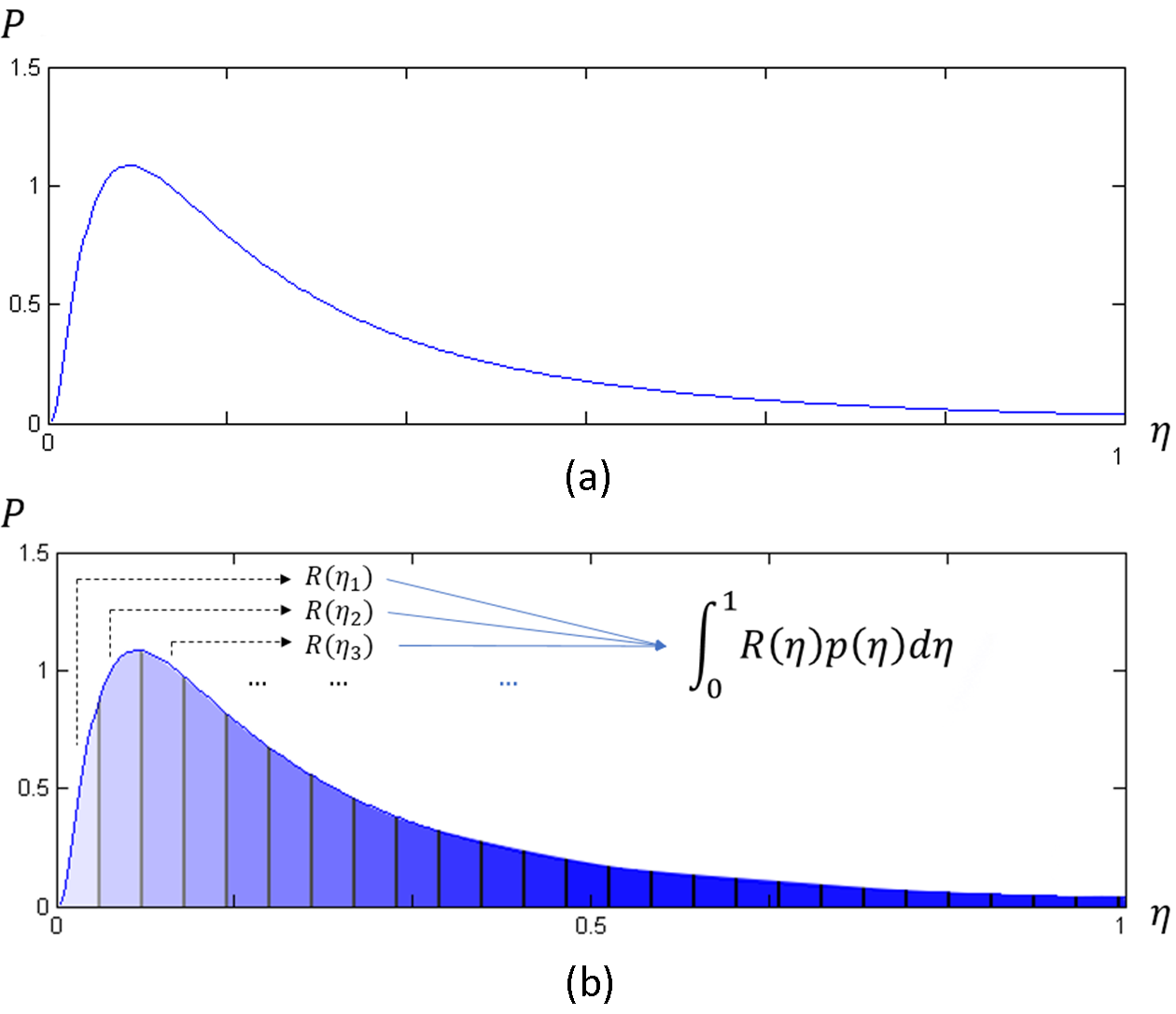}
	\caption{(a) The Probability Distribution of Transmittance Coefficient (PDTC), where $\eta$ is the transmittance, taking a value between [0,1], while P is the probability density function of $\eta$. Here we are showing a plot generated from the log-normal model of the PDTC; (b) Dividing signals into bins according to their respective $\eta$, in the rate-wise integration model}
	\label{fig:PDTC}
\end{figure}

As mentioned in Section I, an important characteristic of a turbulent channel is the time-dependent transmittance, which follows a probability distribution called the PDTC. There have been multiple efforts to accurately characterize the PDTC, and a widely accepted model is the log-normal distribution\cite{distribution,laser} (a plot of which is shown in Fig. \ref{fig:PDTC} (a)):
\begin{equation}
p(\eta)_{\eta_0,\sigma}={1 \over {\sqrt{2\pi \sigma}}\eta}e^{-{{[ln({\eta \over \eta_0})+{1\over 2}{\sigma}^2]^2}\over{2\sigma^2}}}
\end{equation}

\noindent where p is the probability density, $\eta$ the transmittance, and $\eta_0$ and $\sigma$ the mean and variance. The distribution is solely determined by the two parameters $\eta_0$ and $\sigma$, which are inherent to the channel itself. $\eta_0$ is the expected atmospheric transmittance, with a typical value of $10^{-3}$ to $10^{-4}$ (corresponding to 30-40 dB of loss) for a 100km channel, while $\sigma$, typically taking a value between 0 and 1, is determined by the amount of turbulence - the larger the amount of turbulence, the larger the variance. The pair $(\eta_0,\sigma)$ hence contains all information of the PDTC.

Now, we make an important observation: For any given protocol implementation (say, single-photon BB84, or decoy-state BB84), if all experimental parameters in the system except $\eta$ are fixed - i.e. the device parameters including background and dark count rate, detector efficiency, laser intensities, and optical misalignment are all fixed - then the key rate solely depends upon the transmittance $\eta$, and can be written as a single-variable function of $\eta$, i.e. $R(\eta)$. 

To estimate secure key rate of QKD through turbulent channel, the question therefore becomes studying how the function $R(\eta)$ changes, when $\eta$ is a random variable following a probability distribution that we know, the PDTC. 

Here, we will propose two models for $R(\eta)$ under turbulence:

\begin{enumerate}
	\item \textbf{Rate-wise integration model}, $R^{\text{Rate-wise}}$, which is the case where we integrate the rate over PDTC, thus making use of all information of the PDTC. This rate only depends on the rate function and the PDTC, and is independent of what actual threshold we choose.
	\item \textbf{Simplified model}, $R^{\text{Simplified}}(\eta_T)$, which estimates the performance of decoy-state QKD with P-RTS, using post-selection with a threshold $\eta_T$ on channel transmittance. It is a function of the threshold $\eta_T$ that one uses.
\end{enumerate}

Let us first start with the rate-wise integration model. We can begin by considering an ideal case, where we assume that we have complete knowledge of the channel transmittance $\eta$ when each single signal passes through the channel. Moreover, here we discuss the asymptotic case where there is an infinite number of signals sent. Then, it is possible to order all signals from low to high transmittance $\eta$ when they pass through the channel, and divide the signals into bins of $[\eta,\eta+\Delta \eta)$ (which ranges from 0 to 1), as shown in Fig. \ref{fig:PDTC} (b).

Therefore, within the bin $[\eta,\eta+\Delta \eta)$, we can assume that all signals pass through the channel with the same transmittance $\eta$, given that the bin is sufficiently small, i.e. $\Delta \eta \rightarrow 0$. That is, the signals in the same bin can be considered as in a "static channel", and enjoy the same rate formula $R(\eta)$ and security analysis as if $\eta$ is a static constant.

Then, we can calculate the number of secure key bits from each bin, according to their respective $\eta$, and add all bins together. In the limit of $\Delta \eta \rightarrow 0$, this is an integration of $R(\eta)$ over $\eta$, with $p_{\eta_0,\sigma}(\eta)$ being the weight (i.e. the proportion of signals in each bin). We call this model the \textit{"rate-wise integration model"}. Its rate $R^{\text{Rate-wise}}$ satisfies:

\begin{equation}
R^{Rate-wise}=\int_{0}^{1}R(\eta)p_{\eta_0,\sigma}(\eta)d\eta
\end{equation}

$R^{\text{Rate-wise}}$ makes use of all PDTC information from turbulence. Since all bins have either zero or positive rate, using a threshold $\eta_T$ in the rate-wise integration model will always result in either same or smaller rate. i.e.  

\begin{equation}
\begin{aligned}
R^{Rate-wise}(0)&=\int_{0}^{1}R(\eta)p_{\eta_0,\sigma}(\eta)d\eta \\
&\geq \int_{\eta_T}^{1}R(\eta)p_{\eta_0,\sigma}(\eta)d\eta \\
& = R^{\text{Rate-wise}}(\eta_T)
\end{aligned}
\end{equation}

\noindent Hence, from here on if $\eta_T$ is not specified, by $R^{\text{Rate-wise}}$ we will always mean $R^{\text{Rate-wise}}(0)$, which is a constant-value "max possible performance" of key rate that is only dependent on the PDTC of the channel and the experimental device parameters.\\

Now, let us consider applying post-selection to free-space QKD. Using a similar model as in ARTS method (instead of using classical detector voltage V, here we will directly use $\eta$, which is proportional to V). We can set a threshold $\eta_T$ and perform post-selection: we select quantum signals received when transmittance $\eta\geq \eta_T$, and discard all signals received when $\eta<\eta_T$.

Unlike the ideal case of the rate-wise integration model, in reality we do not have infinite resolution from the classical detector, nor do we have an infinite number of signals. In practice, we are post-selecting signals with only two status: "pass" or "fail". To make use of this "pass/fail" information, here we propose a practical model that estimates the rate with only the mean transmittance of the post-selected signals. We name it the \textit{"simplified model"}. First, with no post-selection applied, the rate is:

\begin{equation}
R^{\text{Simplified}}(0)=R(\eta_0)
\end{equation}

\noindent which means that we simply use the mean value of transmittance $\eta_0$ in the channel for all calculations and assume a "static channel", using the same rate formula for a static channel, too. This is, in fact, what has been done in most literature for free-space QKD that don't consider fluctuations due to turbulence, such as in \cite{free2007,Hughes}.

Now, when a threshold is used and post-selection is performed, $R^{\text{Simplified}}$ is written as:

\begin{equation}
R^{\text{Simplified}}(\eta_T)=\int_{\eta_T}^{1}p_{\eta_0,\sigma}(\eta)d\eta \times R(\langle \eta \rangle)
\end{equation}

\noindent here we again treat all post-selected signals as having passed through a "static channel", and use the same rate expression for static case. But the difference is that we use the new mean transmittance among only the post-selected signals, denoted as $\langle \eta \rangle$, as the transmittance of the channel. $\langle \eta \rangle$ satisfies (using expected value formula for a truncated distribution):

\begin{equation}
\langle \eta \rangle={{\int_{\eta_T}^{1}\eta p_{\eta_0,\sigma}(\eta)d\eta}\over{\int_{\eta_T}^{1}p_{\eta_0,\sigma}(\eta)d\eta}}
\end{equation}

When we apply post-selection (like the case with ARTS), in the rate formula for $R^{\text{Simplified}}$, we take into account the loss of signals due to post-selection, and only a portion of $\int_{\eta_T}^{1}p_{\eta_0,\sigma}(\eta)d\eta$ remains. This portion is always no larger than 1, and strictly decreases with $\eta_T$. On the other hand, $\langle \eta \rangle$ is always increasing with $\eta_T$, because we are post-selecting only the signals with higher transmittance. So, just like for the single photon case discussed in Section II.A, we have an optimization problem, where the choice of $\eta_T$ is crucial to the rate we acquire. Using optimal threshold and applying post-selection, as we will later show in the numerical results in the next sections, can dramatically increase the rate over using no post-selection at all.

Therefore, using the simplified model, we can effectively treat the static channel QKD protocol as a "black-box". We enjoy the same rate formula and security analysis as a static channel, while the only difference is that we use a higher $\langle \eta \rangle$ after post-selection as the input, and multiply a reduced portion $\int_{\eta_T}^{1}p_{\eta_0,\sigma}(\eta)d\eta$ to the output.\\

Now, let us compare the performance of the two models. From an information theory perspective, the rate-wise integration model makes use of all possible information on fluctuating transmittance (i.e. the whole PDTC), while the simplified model discards all distribution information and only acknowledges "pass or fail", and keeps only the single mean transmittance after post-selection. Therefore, we expect that the rate-wise integration model, which makes use of the most information, would have a higher rate than the simplified model. We can write the relation as:

\begin{equation}
R^{\text{Rate-wise}} \geq R^{\text{Simplified}}
\end{equation}

This relation suggestions that the Rate-wise integration model is an upper bound for the Simplified model key rate. This result can be shown rigorously by Jensen's Inequality (we include the detailed proof in the Appendix B), under the condition that the rate function $R(\eta)$ is convex. Numerically, we show that (in next section) the rate for single-photon BB84 and decoy-state BB84 are both convex. Therefore, the relation Eq. 7 always holds true.

The next question is, naturally, what is the optimal threshold to choose, such that $R^{\text{Simplified}}$ approaches the upper bound $R^{\text{Rate-wise}}$? Moreover, how closely can it approach the upper bound? We will discuss this optimal threshold in the next section, and show that it is only dependent upon $R(\eta)$ and independent of the PDTC.

\subsection{Optimal Threshold and Near-Tightness of Upper Bound}

In this section, we propose the "Pre-fixed threshold Real Time Selection" (P-RTS) method, and show that the optimal threshold is independent of the PDTC and can be pre-fixed based on experimental parameters only. We also show that with this pre-fixed threshold the simplified model can approach its upper bound very closely.

Here, to describe the key rate function, we have to bring it into the context of an actual protocol model. We will first discuss single-photon BB84, using the Shor-Preskill \cite{Preskill} rate:

\begin{equation}
R=1-2h_2[QBER]
\end{equation}

\noindent here to keep the consistency of notations with following discussions, we will use parameters from Table II (which is also used as the channel model for decoy-state discussion), where detector dark count/background count rate is $Y_0$, basis misalignment is $e_d$, and total system transmittance is $\eta_{sys}=\eta\eta_d$:

\begin{equation}
R_{S-P}=(Y_0+\eta_{sys})\{1-2h_2[e(\eta_{sys})]\}
\end{equation}

\noindent while the single-photon QBER is

\begin{equation}
e(\eta_{sys})={{{1\over 2}Y_0+e_d \eta_{sys}}\over{Y_0+\eta_{sys}}}
\end{equation}

\begin{figure}[t]
	\includegraphics[scale=0.48]{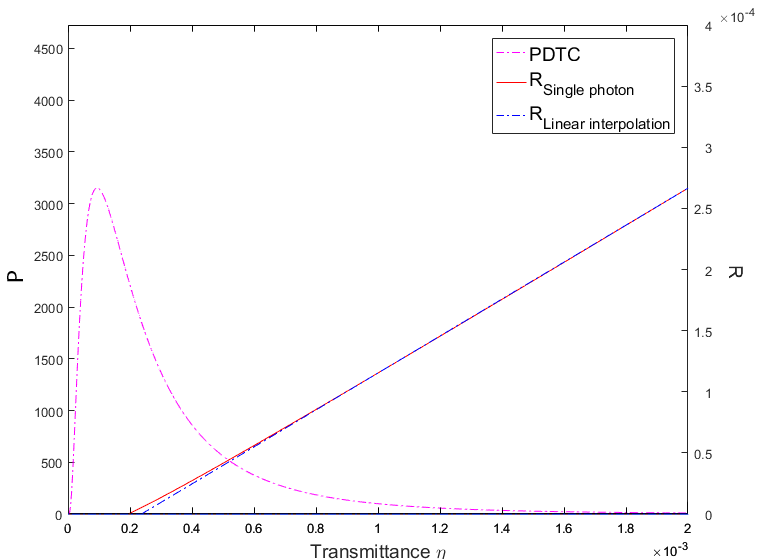}
	\caption{Single-photon rate and PDTC vs Transmittance $\eta$. As can be seen, there is an $\eta_{critical}$ such that $R_{S-P}(\eta)=0$ for all $\eta \leq \eta_{critical}$. For this example, we have plotted the single-photon rate R, using experimental parameters are listed in Table II. We acquire an $\eta_{critical}=0.00020$ for single-photon case. Note that $\eta_{critical}$ is only determined by the experimental parameters of our devices (e.g. dark count rate, and misalignment, and the chosen intensities), and is independent of the actual PDTC. Linear interpolation of the asymptotic $\eta \gg Y_0$ case shows that the function is very close to linear. Here an instance of P, the PDTC function, is also plotted for comparison.}
	\label{fig:critical}
\end{figure}

 A point worth noting is that $R_{S-P}(\eta)$ has the unique property of having an $\eta_{critical}$ such that $R_{S-P}(\eta)=0$ for all $\eta<\eta_{critical}$, and $R_{S-P}(\eta) \geq 0$ for $\eta\geq \eta_{critical}$. This critical position can be expressed as:

\begin{equation}
\eta_{critical}={Y_0 \over \eta_d}{{{1\over 2}-e_{critical}}\over{e_{critical}-e_d}}
\end{equation}

\noindent where $e_{critical}$ is the threshold QBER satisfying 
\begin{equation} 1-2h_2(e_{critical})=0 \end{equation}

\noindent that returns zero rate. For Shor-Preskill rate, this threshold is $e_{critical}=11\%$. More details can be seen in Appendix D.

As can be shown in Fig.\ref{fig:critical}, we plot out the single-photon rate $R_{S-P}(\eta)$, where a sharp turning point $\eta_{critical}$ exists. Moreover, within the $[\eta_{critical},1]$ region, numerical results show that $R(\eta)$ is slightly convex but very close to linear. (For larger $\eta \gg Y_0$, the rate is completely linear. Using this approximation we can make an interpolation of the "linear" part of the rate. As shown in the plot, this linear interpolation is very close to the rate function itself.) 

This can lead to a very interesting result: We showed in Section II.B that $R^{\text{Rate-wise}}$ predicts the maximum possible performance of QKD with threshold post-selection in a turbulence channel. If we choose the threshold $\eta_T=\eta_{critical}$ for the simplified model, we can apply Jensen's Inequality for the truncated $p(\eta)$ distribution within region $[\eta_{critical},1]$, and acquire
\begin{equation}
	R^{\text{Simplified}}(\eta_{critical}) \approx R^{\text{Rate-wise}}(\eta_{critical})
\end{equation}

\noindent given that $R_{S-P}(\eta)$ is very close to linear within the region (but still convex, so $R^{\text{Simplified}}$ is still slightly smaller), since Jensen's Inequality takes equal sign for a linear function. There is also no loss in $R^{\text{Rate-wise}}$ from truncating $[0,\eta_{critical})$, as $R(\eta)=0$ for all $\eta<\eta_{critical}$. 

\begin{equation}
R^{\text{Rate-wise}}(\eta_{critical}) = R^{\text{Rate-wise}}(0)
\end{equation}

\noindent Therefore, $R^{\text{Simplified}}$ can approximately reach the upper bound with $\eta_T=\eta_{critical}$, and the upper bound given by $R^{\text{Rate-wise}}$ is near-tight, due to the near-linearity of $R(\eta)$. A more rigorous proof showing that the optimal threshold for the simplified model is indeed $\eta_{critical}$ can be found in Appendix C.

Also, despite that there is no explicit analytical expression for $\eta_{critical}$, we can show that it depends more heavily on the background/dark count rate (approximately proportional to $Y_0$, if $\eta \ll 1$, and $Y_0 \ll \eta$). Details can be seen in Appendix D.

This result for optimal threshold has two significant implications for using threshold post-selection and applying the simplified model:

\begin{itemize}
	\item  Since $R(\eta)$ is only a function of $\eta$, and not $(\eta_0, \sigma)$, this optimal threshold position $\eta_{critical}$ is only determined by the experimental parameters of the devices (e.g. detector efficiency, dark count rate, misalignment, and Alice's intensities - although here we make an assumption that the misalignment is independent of $\eta$), and thus $\eta_{critical}$ is \textit{\textbf{independent}} of the channel itself and its PDTC. This means that, regardless of the turbulence level, we can use the same threshold to get optimized performance - although the actual amount of performance improvement over not using post-selection \textit{will} be determined by the average loss and the amount of turbulence (i.e. the actual PDTC), as will also be shown in numerical results in the next section. 
	
	\item Given that we choose the optimal threshold and apply P-RTS, not only are we optimizing the rate for the simplified model, but we are also achieving the maximum possible performance for the turbulent channel, even if we make use of all information on transmittance fluctuations. This is because, at $\eta_{critical}$, the max value for $R^{\text{Simplified}}$ can almost reach the upper bound given by the rate-wise integration model - meaning that the upper bound is nearly tight. We will illustrate this point further with numerical results in the next section.
\end{itemize}

The significant implication is that, as long as we know the experimental parameters, we can determine the optimal threshold in advance, without the need to know any information about the channel (such as to measure the turbulence level), and perform post-selection in real time using the fixed threshold.

Therefore, we show that it is possible to perform post-selection on the channel transmittance with a pre-fixed threshold - which we will call "Pre-fixed threshold Real Time Selection" (P-RTS). This is significantly more convenient than protocols which perform optimization of threshold after the experiment is done. It will substantially reduce the amount of data storage requirements in the experiment, since Bob doesn't need to store all data until after experiment for optimization of threshold, and will also save the computational resource since Bob no longer needs to perform optimization of threshold.

\subsection{Numerical Results}

In this section we put the above models into a simulation program for single-photon BB84 in a turbulent channel. We use the experimental parameters from Ref.\cite{free2007}, as listed in Table II. One note is that, though the dark count and stray light contribution is reported to be as high as 1700/s in the paper, because of the gated behavior of the detector and the post-selection, only the counts within a 5.9ns time window (in 100ns period between two pulses, for the 10MHz source used) will affect the result. Therefore, here we take dark count rate as $Y_0=1\times 10^{-5}$ in the simulations.

\begin{table*}[t]
	\caption{Experimental parameters for free-space QKD over an 144km channel in Ref.\cite{free2007}}
	\begin{center}
		\begin{tabular}{ccccccc}			
			dark count rate $Y_0$ & pulse rate & detector efficiency $\eta_d$ & misalignment $e_d$ & error-correction efficiency $f$\\
			\hline
			$1\times 10^{-5}$ (per signal) & 10MHz & 25\% & 3\% & 1.22\\
		\end{tabular}
	\end{center}
\end{table*}

\begin{figure}[t]
	\includegraphics[scale=0.46]{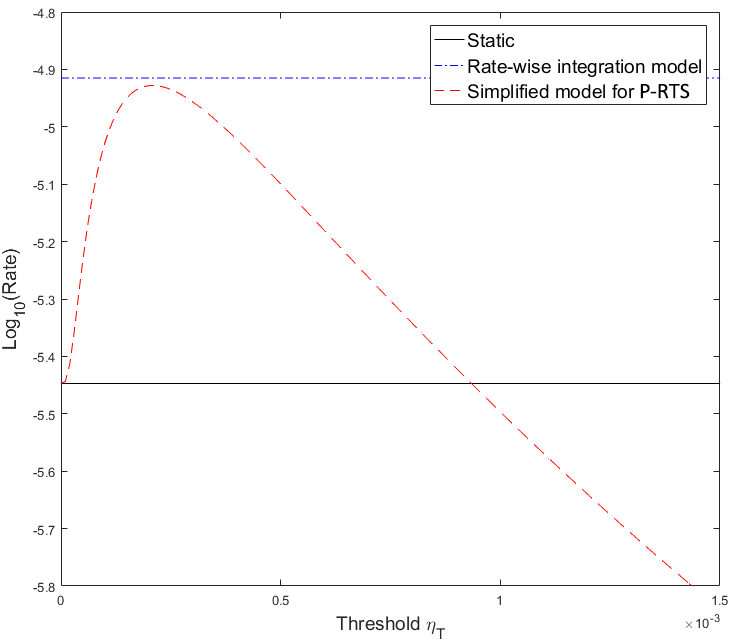}
	\caption{Comparison of the rate-wise integration model and simplified model vs no post-selection (static model), for single-photon case. Here we fix loss=37dB and $\sigma=0.9$, and scan through different threshold $\eta_T$. Experimental parameters are also from Table II. As can be seen, choosing an optimal threshold, which is approximately $\eta_T=0.00020$ here, can allow the simplified model $R^{\text{Simplified}}$ to achieve a rate as much as $97\%$ of - though still lower than - the upper bound given by rate-wise integration model, $R^{\text{Rate-wise}}(0)$.}
	\label{fig:threshold}
\end{figure}

Here, we first take a turbulence level of $\sigma=0.9$, and compare the performance of the two models plus the static model (which is a simplified model with no post-selection, i.e. $R_{S-P}(\eta_0)$) at a fixed loss of 37dB. We plot the results in Fig.\ref{fig:threshold}. As shown in the figure, $R^{\text{Simplified}}(\eta_T)$ first increases with threshold $\eta_T$ (because of post-selecting high-transmittance signals) and then decreases when threshold is further increased (because the decrease in rate due to loss of signals starts to dominate). 

Just as predicted in Section II.C, the simplified model can achieve a very similar performance as the upper bound given by the rate-wise integration model, when the optimal threshold is chosen. For this case, at the optimal threshold $\eta_T=0.00020$, which, as we predicted, is the same as $\eta_{critical}=0.00020$ in Fig.\ref{fig:critical}, we get $R^{\text{Simplified}}=1.18 \times 10^{-5}$, very close to the upper bound $R^{\text{Rate-wise}}=1.22 \times 10^{-5}$ (only by $3\%$ difference - which is due to the rate above $\eta_{critical}$ not perfectly linear), and with dramatic increase in key rate compared with the default static model (using mean transmittance) $R^{\text{Static}}=3.5 \times 10^{-6}$, demonstrating the significant performance gain from using P-RTS in turbulence channel.

\begin{figure}[h]
	\includegraphics[scale=0.48]{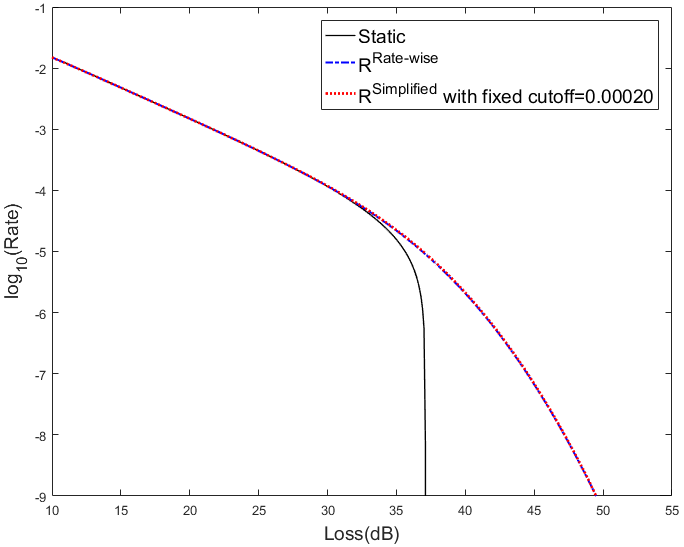}
	\caption{Comparison of the rate-wise integration model, simplified model with optimal threshold, and no post-selection (static model) under $\sigma=0.9$, for the single-photon case. Parameters are from Table II. We can see that the simplified model, with optimized threshold, approaches the rate-wise integration model extremely closely, and both cases have significant improvement in key rate over static (no post-selection) model, especially in high-loss region.}
	\label{fig:turbulence2}
\end{figure}

Furthermore, we compare the rate-wise integration model $R^{\text{Rate-wise}}$, the optimized $R^{\text{Simplified}}(\eta_T)$ with $\eta_T=\eta_{critical}$, and the non-post-selected model (whose rate is equivalent to static model, i.e. $R(\eta_0)$, as in Eq. 4) , by generating the rate vs loss relation for different average loss in the channel. Results can be seen in Fig. \ref{fig:turbulence2}. We see that indeed the rate for simplified model with fixed threshold is extremely close to its upper bound (as suggested in Eq. 13), the rate-wise integration model. Comparing with the static case, we see that the P-RTS method works best for high-loss regions, where post-selection can "salvage" some rate where static case would fail entirely, hence "getting something out of practically nothing". Therefore, one of the major improvements we acquire from using P-RTS in free-space QKD is a dramatically increased maximum tolerant loss (which would mean longer maximum distance).

\section{Decoy-State BB84}

On the other hand, for decoy-state BB84 QKD, we follow decoy-state BB84 QKD theory from Ref. \cite{decoystate_Hwang,decoystate_LMC,decoystate_Wang}, and adopt the notations as in Lo, Ma, and Chen's Paper in 2005 \cite{decoystate_LMC}. Using the GLLP formula \cite{GLLP}, in the asymptotic limit of infinitely many data, we can calculate the secure key rate as:

\begin{equation}
	\begin{aligned}
		R_{GLLP} = q\{-f(E_\mu)Q_\mu h_2(E_\mu)+Q_1[1-h_2(e_1)]\}
	\end{aligned}
\end{equation}

\noindent  where $h_2$ is the binary entropy function, $q={1\over 2}$ or $q\approx 1$ depending on whether efficient BB84 is used, and $f$ is the error-correction efficiency. $Q_\mu$ and $E_\mu$ are the observed Gain and QBER, while $Q_1$ and $e_1$ are the single-photon Gain and QBER contributions estimated using decoy-state. (For a more detailed recapitulation of decoy-state, see Appendix A.1. We have also discussed the channel model that we use for P-RTS in Appendix A.2).

\begin{figure}[h]
	\includegraphics[scale=0.48]{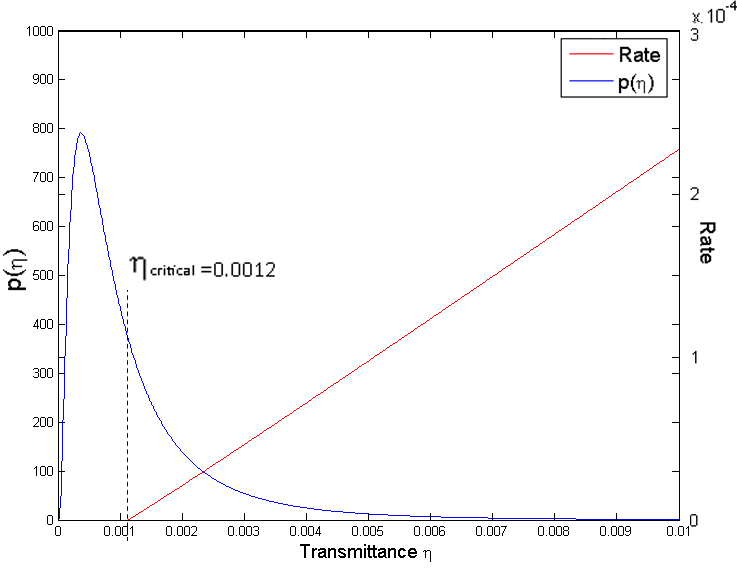}
	\caption{Rate and PDTC vs Transmittance $\eta$ for (asymptotic) decoy-state BB84 with infinite data size. Intensities are $\mu=0.3$, $\nu=0.05$, and experimental parameters are from Table II. As can be seen, there is also an $\eta_{critical}=0.0012$ such that $R_{GLLP}(\eta)=0$ for all $\eta \leq \eta_{critical}$, just like for single photons.}
	\label{fig:critical_decoy}
\end{figure}

Here for free-space decoy-state QKD. We fix the signal and decoy-state intensities as $\mu=0.3$, $\nu=0.05$, and the vacuum state $\omega=0$, and use the vacuum+weak method to estimate single-photon contribution, as in Ma et al.'s 2005 paper \cite{decoypractical} for practical decoy-state QKD. 

Like for the single-photon case, again we generate the rate vs $\eta$ function. As can be observed in Fig. \ref{fig:critical_decoy}, the decoy-state rate function $R_{GLLP}(\eta)$ behaves similarly as the single-photon rate $R_{S-P}(\eta)$, with a critical transmittance $\eta_{critical}$ ($\eta_{critical}=0.0012$ for this parameter set) such that all $\eta$ below it returns zero rate, and a nearly linear rate-transmittance relation for $\eta \geq \eta_{critical}$. Therefore, using the same proof from section II, we can conclude that $\eta_{critical}$ is the optimal (and fixed) threshold for decoy-state BB84 with post-selection too.

\begin{figure}[h]
	\includegraphics[scale=0.48]{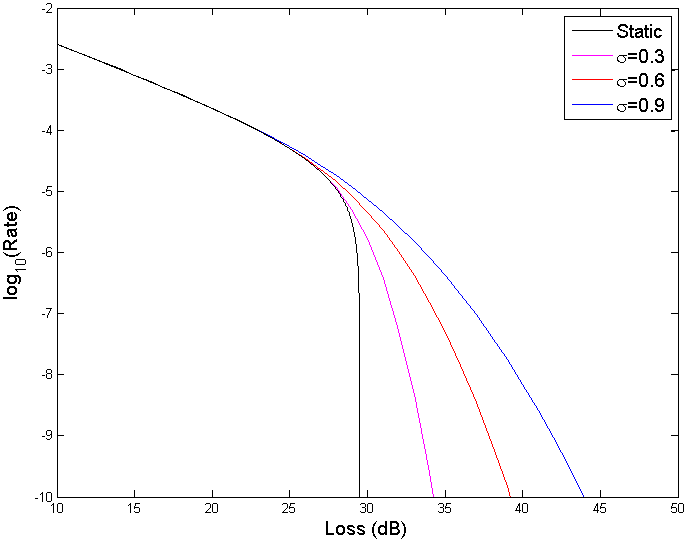}
	\caption{Comparison of the optimized Simplified model vs no post-selection (static model) under different levels of turbulence, for (asymptotic) decoy-state BB84 with infinite data size. Here we use $\sigma=0.3, 0.6, 0.9$ and $\eta_T=0.0012$. Intensities are $\mu=0.3$, $\nu=0.05$, and experimental parameters are from Table II. We see that the improvement in rate from using P-RTS increases with the level of turbulence, and has a significant improvement over static model even under medium-level turbulence of $\sigma=0.6$.}
	\label{fig:turbulence}
\end{figure}

Using the fixed threshold $\eta_T=\eta_{critical}$ to get the optimized rate $R^{\text{Simplified}}(\eta_{critical})$, we generate the rate vs loss relation for different levels or turbulence, as shown in Fig.\ref{fig:turbulence}. As can be seen, the P-RTS method works in the same way with decoy-sate BB84. We can also see that the higher the turbulence level is, the larger performance gain from applying P-RTS will we be able to achieve. As described in Section II.C, the optimal threshold is only determined by the parameters of the equipment, but the actual optimal \textit{performance} is determined by the amount of turbulence present in the channel that we can utilize. As can be seen in the plot, even for a medium-level turbulence of $\sigma=0.6$: for the same loss=29dB, $R^{\text{Simplified}}=8.453 \times 10^{-6}$, a 170\% increase over $R^{\text{Static}}=3.119 \times 10^{-6}$ at loss=29dB. Also, for a minimum rate of $R=10^{-7}$, simplified model has a maximum tolerant loss of 34.4dB, versus 29.5dB for Static Model, with 5.1dB increase in tolerant loss. 

\begin{figure}[h]
	\includegraphics[scale=0.58]{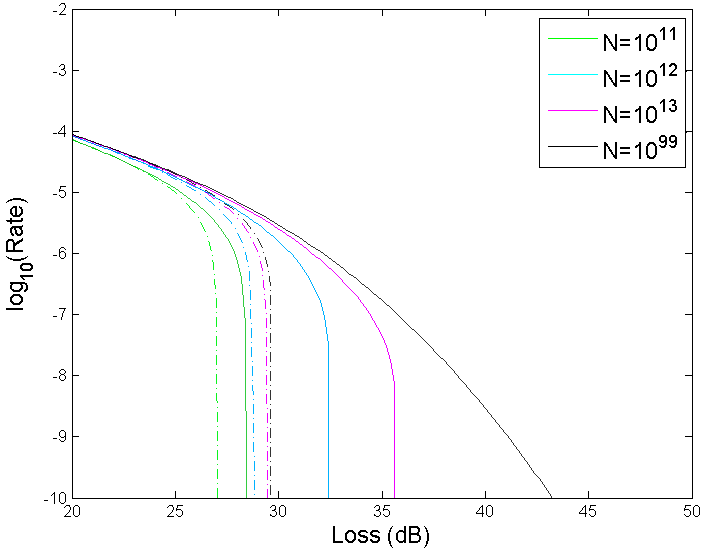}
	\caption{Comparison of the optimized simplified model vs static model, for decoy-state BB84 with finite-size effects. We test different data sizes $N=10^{11},10^{12},10^{13}$, and the near-asymptotic case $N=10^{99}$. Here we use a high turbulence of $\sigma=0.9$. The experimental parameters also follow Table II, and intensities and probabilities used are $\mu=0.31$, $\nu=0.165$, $\omega=2\times 10^{-4}$, $p_{\mu}=0.5$, $p_{\nu}=0.36$, and the probability of sending X basis $q_x=0.75$. The dotted lines are the cases where no post-selection is applied, while the solid lines all have post-selection applied with $\eta_T=0.0012$. We can see from the figure that the improvement in rate from using P-RTS increases with the data size N, and still increases maximum tolerant loss by 3.5dB and 1.4dB when $N=10^{12}$ and $10^{11}$.}
	\label{fig:finite}
\end{figure}

\section{Decoy-State BB84 with Finite-size Effects}

We now turn to the case with finite data size, and apply simplified model and P-RTS to decoy-state BB84 under finite-size effects. We also use simulations to numerically demonstrate the improvements in key rate for finite-size case. The protocol is based on C. Lim et al.'s finite-size decoy-state BB84 paper \cite{finitebb84}, and we have adopted the same channel model as in Ref. \cite{decoypractical}. Here we use the same experimental parameters (including dark count rate, detector efficiency and misalignment) as in Table II, same as the ones used in our previous asymptotic-case simulations. Also, we fix the signal and decoy intensities to $\mu=0.31$, $\nu=0.165$ (in addition to the vacuum intensity $\omega=2\times 10^{-4}$), the probabilities of sending them $p_{\mu}=0.5$, $p_{\nu}=0.36$, and the probability of sending X basis $q_x=0.75$. Unlike in Ref. \cite{finitebb84}, however, we do not scan through the decoy-state intensities and probabilities to perform optimization. Instead, since we only concentrate on high-loss region, we use fixed parameters that are already very close to optimal (while changing them with distance does not provide much improvement in performance). Using intensities that do not change with channel loss also avoids changing the expression for $R_{GLLP}(\eta)$ (which depends on intensities $\mu$, $\nu$), and ensures that $\eta_{critical}$ is independent of the actual loss of the channel.

As described for simplified model, we can use the same "black box" idea, and simply substitute $R_{GLLP}$ for asymptotic BB84 with rate for finite-size BB84. However, one difference from the asymptotic case is that N, the number of signals sent by Alice, matters when calculating the rate, i.e. the rate becomes $R_{Finite-Size}(\eta,N)$ instead of $R_{GLLP}(\eta)$. Then, instead of using Eq. 5 for $R^{\text{Simplified}}$, we use 

\begin{equation}
\begin{aligned}
R^{\text{Simplified}}&=\int_{\eta_T}^{1}p_{\eta_0,\sigma}(\eta)d\eta \\
&\times R_{Finite-Size}(\langle \eta \rangle, N\times \int_{\eta_T}^{1}p_{\eta_0,\sigma}(\eta)d\eta)
\end{aligned}
\end{equation}

\noindent which means that, the post-selection not only affects the overall rate due to the portion of lost signals, but also affects the rate for the \textit{selected} signals, too, since fewer signals than N are used to actually perform the protocol, and higher statistical fluctuations will be present among the selected signals. This means that we need to be even more prudent with post-selection when treating finite-size BB84. 

The numerical results are shown in Fig.\ref{fig:finite}. As can be seen, P-RTS has a similar effect on finite-size BB84 as on the asymptotic case: we gain a significant advantage in the high-loss region, and have an improved maximum tolerant loss, when a minimum acceptable Rate is required. For instance, at $\sigma=0.9$ and for a minimum $R=10^{-7}$, the maximum loss increases by 1.4dB, 3.5dB, 5.2dB, and 6.2dB, respectively, for the cases with $N=10^{11}, 10^{12}, 10^{13}$, and near-asymptotic case, while not much improvement can be gained from P-RTS with N smaller than $10^{10}$. As shown, the improvement increases with the size of N (which is understandable, since the smaller N is, the more sensitive the rate will be to post-selection - because we are cutting off a portion from the already-insufficient number of signals and further aggravating the statistical fluctuations - while for the asymptotic case, for instance, the performance of selected signals does not depend upon how big is the selected portion of signals, and the only negative effect post-selection has is the lost portion of signals). For a free-space QKD system with 100MHz repetition rate, $N=10^{11}$ would require about 17 minutes of communication.

\section{Conclusion}

In this paper we have proposed a post-selection method with prefixed threshold for QKD through turbulent channel, and have also proposed a universal framework for determining the optimal threshold beforehand and predicting the maximum possible performance. By choosing the threshold in advance, we can perform post-selection in real time regardless of the channel condition. This real-time post-selection also provides an additional benefit of reducing the amount of data that is required to be stored in the detector system on Bob's side. We also performed simulations to show the method's effectiveness in not only single-photon BB84, but also practical decoy-state QKD in both the asymptotic case and the case with finite-size effects.

This method is especially effective for regions of high turbulence and high loss, and can even "salvage something out of nothing", when the secure key rate could have been zero without P-RTS method. In order to sample the real-time transmittance condition, the P-RTS method can use only an additional classical channel for each quantum channel, which would be easily implemented (or may even be already implemented as a beacon laser is often required for alignment in free-space QKD). Moreover, since our results only depend on post-selection of $\eta$, in essence our method is even possible without an additional classical channel, such as in Erven et al.'s SNRF setup \cite{SNRF} (which samples transmittance by observing quantum signal count rate). The thresholding, on the other hand, is purely implemented in post-processing, therefore does not require any additional resource, and could be readily deployed into existent infrastructure, and gain a ready increase in secure key rate performance over existing implementation for free-space QKD. 

\section{Acknowledgements}

This work was supported by the Natural Sciences and Engineering Research Council of Canada (NSERC), U.S. Office of Naval Research (ONR), and China 1000 Young Talents Program. We thank the collaborators B Qi and G Siopsis on helpful discussions and collaborative efforts, K McBryde and S Hammel for kindly providing atmospheric data and for the helpful discussions, JP Bourgoin, P Chaiwongkhot, T Jennewein, A Hill and P Kwiat for illuminating discussions on turbulence, and Sheng-Kai Liao and Cheng-Zhi Peng for helpful discussions on free-space QKD.



\appendix

	\section{Decoy-State BB84 Rate Function}
	
	\subsection{Standard Channel Model}
	
	Here we present a brief recapitulation of the decoy-state BB84 model we used. We follow the notations as in Lo, Ma, and Chen's Paper in 2005 \cite{decoystate_LMC}.
	
	Alice uses a WCP source at intensity $\mu$, which sends pulses with a Poissonian photon number distribution: $P_i={\mu^i \over i!}e^{-\mu}$. We will first consider using the standard channel model (as in the original paper Ref.\cite{decoystate_LMC}), where for each i-photon pulse $\ket{i}$, the transmittance, yield $Y_i$, gain $Q_i$, and QBER $e_i$ are:
	\begin{equation}
	\begin{aligned}
	\eta_i&=1-(1-\eta)^i\\
	Y_i&\approx Y_0+\eta_i=Y_0+1-(1-\eta)^i\\
	Q_i&=Y_i{\mu^i \over i!}e^{-\mu}\\
	e_i&={{e_0Y_0+e_d \eta _i}\over Y_i}\\
	\end{aligned}
	\end{equation}
	
	\noindent where $Y_0$, $e_d$ are the dark count rate and misalignment, respectively, and $e_0={1\over 2}$. The overall Gain $Q_{\mu}$ and QBER ${E_{\mu}}$ for this intensity $\mu$ are:
	
	\begin{equation}
	\begin{aligned}
	Q_{\mu}&=\sum_{i=0}^{\infty}Y_i{\mu^i \over i!} e^{-\mu}=\sum_{i=0}^{\infty}Y_iP_i \\
	E_\mu &= {1 \over Q_\mu} \sum_{i=0}^{\infty}e_i Y_i{\mu^i \over i!} e^{-\mu}={1 \over Q_\mu} \sum_{i=0}^{\infty}e_iY_iP_i
	\end{aligned}
	\end{equation}
	
	\noindent where $Q_{\mu}$ and $E_{\mu}$ are simulated here for rate estimation using known channel transmittance $\eta$, while in experiment they will be measured observables. 
	
	For this standard channel model, we assume that the photon number distribution after passing through the channel would still be Gaussian. Using decoy-state technique to combine $Q_{\mu}$ and $E_{\mu}$ for different intensities, we can estimate the single-photon contributions $Q_1$ and $e_1$. The achievable secure key rate is at least
	
	\begin{equation}
	\begin{aligned}
	R_{GLLP} = q\{-f(E_\mu)Q_\mu h_2(E_\mu)+Q_1[1-h_2(e_1)]\}
	\end{aligned}
	\end{equation}
	
	\noindent as given by the GLLP formula\cite{GLLP}, where $h_2$ is the binary entropy function, $q={1\over 2}$ or $q\approx 1$ depending on whether efficient BB84 is used, and $f$ is the error-correction efficiency.
	
	\subsection{Channel Model after Post-Selection}
	
	However, one thing worth noting is that although photon number distribution is Gaussian after the signals pass through the standard channel model, it is no longer necessarily so if we perform post-selection, in which case the photon number distribution might change, and thus the decoy-state key rate form in Eq. A3 (which depends on a Gaussian distribution model) might no longer be adequate.
	
	To show that this will not be a concern for us, we will explicitly discuss how the post-selection from P-RTS will affect the yield for each photon number. From Eq. A1, before post-selection, the yield for pulses with a given photon number $i$ is
	
	\begin{equation}
	\begin{aligned}
	Y_i(\eta)=Y_0+1-(1-\eta)^i
	\end{aligned}
	\end{equation}
	
	For simplified model, among the post-selected signals, we have replaced $\eta$ in Eq. A4 with 
	\begin{equation}
	\begin{aligned}	
	\langle \eta \rangle={{\int_{\eta_T}^{1}\eta p_{\eta_0,\sigma}(\eta)d\eta}\over{\int_{\eta_T}^{1}p_{\eta_0,\sigma}(\eta)d\eta}}
	\end{aligned}
	\end{equation}
	
	\noindent thus the yield for i-photon pulse is assumed to be:
	
	\begin{equation}
	\begin{aligned}
	Y_i^{Simplified}(\eta_T)=Y_0+1-(1-\langle \eta \rangle)^i
	\end{aligned}
	\end{equation}
	
	\noindent in which case, we are simply replacing the $\eta$ with a higher expected value $\langle \eta \rangle$, but the expression is in the same form as Eq. A4, and the received photon number distribution is still Gaussian. (Hence decoy-state analysis and key rate expression still hold).
	
	However, if we consider the more realistic case, post-selection might have a different effect on pulses with different photon number $i$. Therefore, to estimate the yield for each photon number, and analyze the photon number distribution after the channel and the post-selection, we should group up pulses with the same given photon number, and calculate the expected value of the yield for each given $i$. We can call this the "pulse-wise integration" model.
	
	\begin{equation}
	\begin{aligned}
	Y_i^{Pulse-wise}(\eta_T)={{\int_{\eta_T}^{1}Y_i(\eta) p_{\eta_0,\sigma}(\eta)d\eta}\over{\int_{\eta_T}^{1}p_{\eta_0,\sigma}(\eta)d\eta}}=\langle Y_i(\eta) \rangle
	\end{aligned}
	\end{equation}
	
	\noindent and the Gain $Q_{\mu}$ and QBER ${E_{\mu}}$ would become:
	
	\begin{equation}
	\begin{aligned}
	Q_{\mu}&=\sum_{i=0}^{\infty}\langle Y_i(\eta) \rangle{\mu^i \over i!} e^{-\mu}=\sum_{i=0}^{\infty}\langle Y_i(\eta) \rangle P_i \\
	E_\mu &= {1 \over Q_\mu} \sum_{i=0}^{\infty}e_i \langle Y_i(\eta) \rangle{\mu^i \over i!} e^{-\mu}={1 \over Q_\mu} \sum_{i=0}^{\infty}e_i \langle Y_i(\eta) \rangle P_i
	\end{aligned}
	\end{equation}
	
	In the case of this "pulse-wise integration" model, $Q_\mu$ and $E_\mu$ can no longer be considered as from a Gaussian distribution with intensity $\eta\mu$, which is seemingly warning us that the decoy-state analysis might not hold true anymore. However, here we make the observation that for $i=0$, trivially, 
	\begin{equation}
	\begin{aligned}
	Y_0^{Simplified}=Y_0=Y_0^{Pulse-wise}
	\end{aligned}
	\end{equation}
	
	\noindent and for $i=1$, the yield is a linear function of $\eta$, hence
	
	\begin{equation}
	\begin{aligned}
	Y_1^{Simplified}=Y_0+\langle \eta \rangle = \langle(Y_0+\eta)\rangle=Y_1^{Pulse-wise}
	\end{aligned}
	\end{equation}
	
	While for all multi-photon cases where $i\geq 2$, the function 
	
	\begin{equation}
	\begin{aligned}
	Y_i(\eta)=Y_0+1-(1-\eta)^i
	\end{aligned}
	\end{equation}
	
	is a strictly concave function on the domain $[0,1]$. Therefore, from Jensen's Inequality, the expected value of a concave function is strictly smaller than the function ($Y_i$) of the expected value, i.e.
	
	\begin{equation}
	\begin{aligned}
	Y_i^{Pulse-wise}=\langle Y_i(\eta)\rangle < Y_i(\langle \eta \rangle) = Y_i^{Simplified}, i\geq 2
	\end{aligned}
	\end{equation}
	
	This means that, with the simplified model, with the Gaussian photon number distribution assumption and the standard decoy-state key rate analysis, we are correctly estimating the vacuum and single-photon contributions, but always \textit{over-estimating} the multi-photon contributions. This will in fact result in an \textit{under-estimated} key rate for simplified model than the realistic case (yield-wise integration model). Therefore, we make the "validity argument" here that, despite post-selection will result in a non-Gaussian photon number distribution, by using simplified model and the same decoy-state analysis, we will never incorrectly over-estimate the key rate, and can be confident in the improvement in performance from using P-RTS.

	\section{Proof of Rate-Wise Integration Model as Upper Bound}
	
	To better compare the models, let us first simply the notations, and define $\langle f(\eta)\rangle$ operator as taking the expected value of $f(\eta)$ over $p_{\eta_0,\sigma}(\eta)$ (in the case of using post-selection, the distribution is truncated, and will be normalized by dividing by $\int_{\eta_T}^{1}p_{\eta_0,\sigma}(\eta)d\eta$). The expected value $\langle f(\eta) \rangle$ can be expressed as:
	
	\begin{equation}
	\langle f(\eta) \rangle={{\int_{\eta_T}^{1}f(\eta) p_{\eta_0,\sigma}(\eta)d\eta}\over{\int_{\eta_T}^{1}p_{\eta_0,\sigma}(\eta)d\eta}}
	\end{equation}
	
	Then, we can easily see that, mathematically, the two models we proposed so far, the rate-wise integration model and the simplified model, are only different in that they apply the "expected value" operator at different levels of the function. We can simply write $R^{\text{Rate-wise}}$ and $R^{\text{Simplified}}$ as:
	
	\begin{equation}
	\begin{aligned}
	R^{\text{Rate-wise}} (\eta_T)&=\langle R(\eta) \rangle\\
	R^{\text{Simplified}} (\eta_T)&=R(\langle \eta \rangle)
	\end{aligned}
	\end{equation}
	
	Now, we introduce the \textit{Jensen's Inequality}: \\
	
	\textit {For a random variable X following a probability distribution p(X), and for any given convex function f(x), we always have}
	\begin{equation}
	\langle f(X) \rangle \geq f(\langle X \rangle)
	\end{equation}
	
	\noindent the equal sign is taken when the function $f(x)$ is linear.
	
	For decoy-state BB84, $R_{GLLP}(\eta)$ is a convex (and increasing) function of $\eta$, therefore we have $\langle R(\eta) \rangle \geq R(\langle \eta \rangle)$, i.e. 
	
	\begin{equation}
	R^{\text{Rate-wise}}(0) \geq R^{\text{Simplified}}(0)
	\end{equation}
	
	\noindent This holds true even after a threshold is applied, too, since we can simply replace the distribution $p(\eta)$ with the truncated distribution on domain $[\eta_T,1]$, and normalize it by dividing by the constant $\int_{\eta_T}^{1}p_{\eta_0,\sigma}(\eta)d\eta$. Since $R(\eta)$ is non-concave on all sections of $[0,1]$, the Jensen's Inequality always holds true, regardless of the threshold. i.e.
	
	\begin{equation}
	R^{\text{Rate-wise}}(0) \geq R^{\text{Rate-wise}}(\eta_T) \geq R^{\text{Simplified}}(\eta_T)
	\end{equation}
	
	\noindent here we also include Eq. 3's result that $R^{\text{Rate-wise}}(\eta_T)$ is non-increasing with $\eta_T$.
	
	Therefore, we see that $R^{\text{Rate-wise}}$ serves as an upper bound for the possible rate in a turbulent channel, as it is the maximum achievable rate when we know all transmittance information and make use of the entire PDTC. simplified model always has no higher rate than this upper bound. This means that, when we use $R^{\text{Simplified}}$ to calculate the rate, we \textit{never overestimate} the performance of the protocol. When we demonstrate the improvements we gain by using P-RTS in decoy-state BB84, the actual possible rate will be even higher, thus the validity argument for the usage of the simplified model in estimating the rate. 
	
	\section{Proof of Optimality of Critical Transmittance as Threshold for Simplified Model}
	
	Following the argument in Section II.C, here we give a rigorous proof that $\eta_T=\eta_{critical}$ is indeed the optimal threshold for the simplified model, given that $R_{S-P}(\eta)$ (and similarly for $R_{GLLP}(\eta)$) is nearly linear. For simplified model, we showed that
	
	\begin{equation}
	R^{\text{Simplified}}(\eta_T)=\int_{\eta_T}^{1}p_{\eta_0,\sigma}(\eta)d\eta \times R_{S-P}(\langle \eta \rangle)
	\end{equation}
	
	\noindent where $\langle \eta \rangle$ satisfies:
	
	\begin{equation}
	\langle \eta \rangle={{\int_{\eta_T}^{1}\eta p_{\eta_0,\sigma}(\eta)d\eta}\over{\int_{\eta_T}^{1}p_{\eta_0,\sigma}(\eta)d\eta}}
	\end{equation}
	
	\noindent Then, using the Leibniz Integration Rule, and taking derivative with respect to $\eta_T$ (here we omit the subscript of $\eta_0,\sigma$ for the PDTC, and $S-P$ (or $GLLP$) for the rate), we have
	
	\begin{equation}
	\begin{aligned}
	{d \over d\eta_T}{\langle \eta \rangle}={{p(\eta_T)}\over{\int_{\eta_T}^{1}p(\eta)d\eta}}({\langle \eta \rangle} - \eta_T)
	\end{aligned}
	\end{equation}
	
	\noindent using the chain rule, 
	
	\begin{equation}
	\begin{aligned}
	{d \over d\eta_T}R(\langle \eta \rangle) &= {dR(\eta) \over {d\langle \eta \rangle}} {{d\langle \eta \rangle}\over d\eta_T}\\
	&= R'(\langle \eta \rangle) {{p(\eta_T)}\over{\int_{\eta_T}^{1}p(\eta)d\eta}}({\langle \eta \rangle} - \eta_T)\\
	\end{aligned}
	\end{equation}
	
	\noindent Maximizing $R^{\text{Simplified}}$ requires that 
	\begin{equation}
	{d \over d\eta_T}R^{\text{Simplified}}(\eta_T)=0
	\end{equation}

	\noindent expanding the derivative using Eq. C1 gives us
	\begin{equation}
	\begin{aligned}
	&{d \over d\eta_T}R^{\text{Simplified}}(\eta_T)\\
	 &= \left( \int_{\eta_T}^{1}p(\eta)d\eta \right) \times {d \over d\eta_T}R(\langle \eta \rangle) - p(\eta_T)R(\langle \eta \rangle) \\
	&= R'(\langle \eta \rangle) {p(\eta_T)}({\langle \eta \rangle} - \eta_T) - p(\eta_T)R(\langle \eta \rangle) \\
	&= p(\eta_T)[({\langle \eta \rangle} - \eta_T) R'(\langle \eta \rangle) - R(\langle \eta \rangle)]
	\end{aligned}
	\end{equation}
	
	\noindent Therefore, the optimal threshold requires that 
	
	\begin{equation}
	({\langle \eta \rangle} - \eta_T) R'(\langle \eta \rangle) = R(\langle \eta \rangle)
	\end{equation}\\
	
	When $R(\eta)$ is a linear function on the domain $[\eta_{critical}, 1]$ and $R(\eta_{critical})=0$, there is
	
	\begin{equation}
	R'(\langle \eta \rangle) = {{R(\langle \eta \rangle) - R(\eta_T)}\over{({\langle \eta \rangle} - \eta_T)}}
	\end{equation}
	
	\noindent combined with Eq. C7, we have
	
	\begin{equation}
	R(\eta_T)=0
	\end{equation}
	
	\noindent for $\eta \in [\eta_{critical}, 1]$, there is one and only one point satisfying $R(\eta_T)=0$, that is 
	\begin{equation}
	\eta_T=\eta_{critical}
	\end{equation}
	
	\noindent For $\eta \in [0, \eta_{critical})$, on the other hand,
	
	\begin{equation}
	R(\langle \eta \rangle) - R(\eta_T) < R'(\langle \eta \rangle) ({\langle \eta \rangle} - \eta_T) = R(\langle \eta \rangle)
	\end{equation}
	
	\noindent which becomes
	
	\begin{equation}
	R(\eta_T) > 0
	\end{equation}
	
	\noindent but $R(\eta)=0$ for all $\eta \leq \eta_{critical}$, so no $\eta_T\in [0, \eta_{critical})$ satisfies the zero derivative requirement. Which means that, when $R_{GLLP}(\eta)$ is near linear on $[\eta_{critical}, 1]$, we have
	
	\begin{equation}
	\eta_T=\eta_{critical}
	\end{equation}
	
	\noindent as the one and only optimal threshold for $R^{\text{Simplified}}$.\\
	
	Additionally, if we do not ignore the convexity of $R(\eta)$, consider the tangent line for $R(\eta)$ at $\langle \eta \rangle$, since $R(\eta)$ is a convex function of $\eta$,
	
	\begin{equation}
	({\langle \eta \rangle} - \eta_T) R'(\langle \eta \rangle) > R(\langle \eta \rangle) - R(\eta_T)
	\end{equation}
	
	\noindent optimal threshold requires that 
	
	\begin{equation}
	R(\langle \eta \rangle) > R(\langle \eta \rangle) - R(\eta_T)
	\end{equation}
	
	\noindent i.e. $R(\eta_T) > 0$, which means that the optimal threshold position will be shifted rightward from $\eta_{critical}$, the actual amount of shift depends on how much $R$ deviates from linearity (in numerical simulations, we see that since $R(\eta)$ is very close to linear, this shift is very small). Also, although $R^{\text{Rate-wise}}$ is not affected for a threshold no larger than $\eta_{critical}$, using a threshold larger than $\eta_{critical}$ will cause $R^{\text{Rate-wise}}$ to decrease, since "bins" with positive rate are discarded. Therefore, the maximum point for $R^{\text{Simplified}}$ is no longer the maximum $R^{\text{Rate-wise}}$, but slightly smaller than it. This also explains why in the numerical results, the optimal $R^{\text{Simplified}}$ is always slightly lower than upper bound, due to non-linearity of $R(\eta)$.
	
	Also, a small note is that, the Jensen's Inequality asks the function to be differentiable at every point, while the turning point of $R$ at $\eta_{critical}$ is a sharp point. To address this, we can construct another $R_2$ with an infinitesimally small yet smooth "turn" at $\eta_{critical}$ to replace the sharp point, but as the "turn" is infinitely small, integrating $R$ and $R_2$ over any region will yield infinitely close results. Therefore the turning point's structure does not affect the above results.
	
	\section{Analytical Expression for Optimal Threshold}
	
	\subsection{Single-Photon Case}
	
	Let $\eta_{sys}=\eta\times \eta_d$. The single photon Shor-Preskill rate is
	
	\begin{equation}
	R_{S-P}=(Y_0+\eta_{sys})\{1-2h_2[e(\eta_{sys})]\}
	\end{equation}
	
	where the single-photon QBER is
	
	\begin{equation}
	e(\eta_{sys})={{{1\over 2}Y_0+e_d \eta_{sys}}\over{Y_0+\eta_{sys}}}
	\end{equation}
	
	For the rate to be zero, we require:
	
	\begin{equation}
	R_{S-P}=0
	\end{equation}
	
	hence
	
	\begin{equation}
	1-2h_2[e(\eta_{sys})]=0
	\end{equation}
	
	or, $h_2[e(\eta_{sys})]={1\over 2}$. This numerically corresponds to $e(\eta_{sys})=11\%=e_{critical}$ (which is the QBER threshold for Shor-Preskill rate). Therefore, substituting into Eq. D2, we have
	
	\begin{equation}
	\eta_{sys}={{{1\over 2}-e_{critical}}\over{e_{critical}-e_d}}Y_0
	\end{equation}
	
	expressing it in channel transmittance $\eta$
	
	\begin{equation}
	\eta_{critical}={Y_0 \over \eta_d}{{{1\over 2}-e_{critical}}\over{e_{critical}-e_d}}
	\end{equation}
	
	\noindent or, if we substitute $\eta_{critical}=11\%$ into the equation, we have
	
	\begin{equation}
	\eta_{critical}={Y_0 \over \eta_d}{0.39\over{0.11-e_d}}
	\end{equation}
	
	This is the analytical expression for the critical transmittance for the single-photon case. Also, we can see that the critical transmittance is proportional to the background count (i.e. noise) in the system. i.e.
	
	\begin{equation}
	\eta_{critical} \propto{Y_0 \over \eta_d}
	\end{equation}
	
	\subsection{Decoy-State BB84}
	
	Consider the asymptotic case of decoy-state BB84, with infinite number of decoys (i.e. the only significant intensity is the signal intensity $\mu$). Using the GLLP rate,
	
	\begin{equation}
	\begin{aligned}
	R_{GLLP} = q\{-fQ_\mu h_2(E_\mu)+Q_1[1-h_2(e_1)]\}
	\end{aligned}
	\end{equation}
	
	\noindent we would like to find $\eta_{critical}$ such that 
	\begin{equation}
	R(\eta_{critical})=0
	\end{equation}
	
	\noindent hence
	\begin{equation}
	fQ_\mu h_2(E_\mu)=Q_1[1-h_2(e_1)]
	\end{equation}
	
	\noindent or
	
	\begin{equation}
	h_2(e_1)+f{Q_\mu \over Q_{1}} h_2(E_\mu)=1
	\end{equation}
	
	Let $\eta_{sys}=\eta\times \eta_d$, the observables and single-photon contributions can be written as:
	
	\begin{equation}
	\begin{aligned}
	Q_\mu &= Y_0+1-exp(-\mu\eta_{sys})\\
	E_\mu &= {{{1\over 2}Y_0+e_d(1-exp(-\mu\eta_{sys}))}\over{Y_0+1-exp(-\mu\eta_{sys})}}\\
	Q_1 &=\mu exp(-\mu)(Y_0+\eta_{sys})\\
	e_1 &={{{1\over 2}Y_0+e_d\eta_{sys}}\over{Y_0+\eta_{sys}}}
	\end{aligned}
	\end{equation}
	
	Now, if $\eta \ll 1$, we can use the approximation $1-exp(-\mu\eta_{sys})=\mu\eta_{sys}$. If the dark/background count rate $Y_0$ also satisfies $Y_0 \ll \eta_{sys}$ (which is a reasonable approximation, since with parameters in Table II, $Y_0$ is at the order of $10^{-5}$, while $\eta_d \eta_{critical}$ is at the order of $10^{-3}$), we can write
	
	\begin{equation}
	\begin{aligned}
	{Q_\mu \over Q_1 }&\approx {{Y_0+\mu\eta_{sys}}\over{\mu exp(-\mu)(Y_0+\eta_{sys})}} \approx exp(\mu)\\
	e_1 &\approx {1\over 2}{Y_0 \over \eta_{sys}}+e_d \\
	E_\mu &\approx {1\over {2\mu}}{Y_0 \over \eta_{sys}}+e_d
	\end{aligned}
	\end{equation}
	
	\noindent substituting back into Eq. D12, and defining
	
	\begin{equation}
	x={Y_0 \over \eta_{sys}}={Y_0\over{\eta_d\eta}}
	\end{equation}
	
	\noindent we can have
	
	\begin{equation}
		h_2({1\over 2}x+e_d)+fe^\mu h_2({1\over {2\mu}}x+e_d)=1
	\end{equation}
		
	\noindent this is a function that is only determined by $e_d$ and $\mu$. We can write its solution for x as
	
	\begin{equation}
		x_{critical}=\mathcal{F}(e_d,\mu)
	\end{equation}
	
	Then the critical transmittance (i.e. optimal threshold position) can be written as
	
	\begin{equation}
		\eta_{critical}={{Y_0}\over{\eta_d}} [{1\over\mathcal{F}(e_d,\mu)}]
	\end{equation}
	
	\noindent where $\mathcal{F}(e_d,\mu)$ does not have an explicit analytical expression, because $h_2$ function cannot be analytically expanded. (One can, however, numerically use linear fit to expand $h_2$, if given the approximate range of the experimental parameters $e_d$ and $\mu$). The important observation here, however, is that for the decoy-state case, we can still have:
	
	\begin{equation}
	\eta_{critical} \propto{Y_0 \over \eta_d}
	\end{equation}
	
	\noindent which points out that the critical threshold is directly proportional to the dark (or background) count rate of the experimental devices, and inversely proportional to the detector efficiency.
	
	\section{PDTC parameters}
	
	In our simulations, we have fixed several typical values for $\sigma$ for free-space QKD, corresponding to the case of weak-to-medium level turbulence, and have considered the PDTC to be a fixed distribution for a given $\sigma$ regardless of the channel loss. In reality, though, $\sigma$ is distance-dependent, too. A commonly used estimation for $\sigma$ is the "Rytov Approximation"\cite{rytov}
	
	\begin{equation}
	\sigma^2 = 1.23 C_n^2 k^{7/6} L^{11/6}
	\end{equation}
	
	\noindent which relates $\sigma$ both to the distance $L$ and the refractive index structure constant $C_n^2$ (which is determined by atmospheric conditions). 
	
	Also, with simulation software such as MODTRAN\cite{modtran}, it is possible to simulate the relationship between $\eta_0$ and $L$ for a given free-space channel. Therefore, one necessary next step would also be to estimate performance for cases with realistic values for $\eta_0$ and $\sigma$, both from literature and from simulations, as well as to study the possible correlation $\sigma$ and $\eta_0$ (both related to $L$) in simulations.

\end{document}